# Roles of Structural Coordination and Strain Orientation in the Phase Stability of Ferroelectric HfO$_2$

Adedamola D. Aladese and Xiao Shen[*]
Department of Physics and Materials Science, The University of Memphis,
Memphis, Tennessee 38152, USA

**Abstract**

Phase stabilization continues to be a critical issue in hafnium oxide (HfO$_2$) due to the interdependence of various contributing factors. Using first-principles calculations, we analyze the effects of strain and doping on stabilizing the ferroelectric phase. We found that combining Y-doping, O-vacancy, and compressive biaxial strain, particularly in the (111) orientation, offers an optimal pathway for stabilizing the ferroelectric phase of HfO$_2$. Analysis of structural coordination reveals how compressive strain affects phase competition. Crystallography analysis provides insights into the advantage of the (111) strain orientation compared to the (001) orientation. The impact of dopants is discussed in the context of these findings.

## I. INTRODUCTION

The miniaturization of silicon technology has driven the performance of electronic devices in the past decades. However, the challenges with Moore's scaling law and the designs for next-generation computing architectures have led to increased research into alternative materials. Although hafnium oxide (HfO$_2$) has been adopted in the silicon industry for high-k applications,[1] the discovery of ferroelectricity in this material is a significant new opportunity in the semiconductor industry.[2] Unlike traditional perovskites whose ferroelectricity disappears at lower dimensions, HfO$_2$ showed robust ferroelectricity even in reduced thickness.[3,4] Moreover, the easy integration with CMOS technology makes HfO$_2$-based ferroelectric memories strong candidates for future electronics.[5–9]

HfO$_2$ has various polymorphs, including the most stable monoclinic (P2$_1$/c), tetragonal (P4$_2$/nmc), cubic (Fm$\bar{3}$m), and orthorhombic (Pca2$_1$, Pbca, Pmn2$_1$, etc.) phases.[10] The observed ferroelectric behavior in HfO$_2$ thin films has been attributed to the formation of the polar orthorhombic Pca2$_1$ phase, arising in the presence of extrinsic factors such as strain, electric field, oxygen vacancies, and dopants.[4,11–14] Stabilizing the metastable Pca2$_1$ phase over the monoclinic phase is challenging owing to the interplay of several mechanisms in phase competition.[15–17] Batra *et al.* proposed that the combination of biaxial strain and electric field is effective in stabilizing the Pca2$_1$ phase instead of a single factor.[11] Also, trivalent dopants have been shown to facilitate phase transition and stability by compensating for oxygen vacancies.[12,18] Using a strain engineering approach, Liu *et al.* showed that biaxial strain in the (111) orientation in pure HfO$_2$ makes the Pca2$_1$ phase the most stable.[16] On the contrary, Zhang *et al.* showed that the monoclinic phase remains the most stable in the (111) orientation.[19] Overall, the nature of ferroelectricity in HfO$_2$ is not yet fully understood.

---

[*] Email: xshen1@memphis.edu

This paper presents a first-principles study of the effects of different strain orientations on phase stability and the roles of doping. We find that although both (001) and (111) biaxial compressive strains can stabilize the ferroelectric phase, the (111) orientation is more effective. Co-doping with yttrium atoms and oxygen vacancies further enhances this effect. We attribute the beneficial impact of compressive strain to the flexibility around the tri-coordinated oxygen atoms in the ferroelectric phase. The effectiveness of the (111) strain orientation arises from its higher atomic density. Co-doping with yttrium and oxygen vacancies enhances the flexibility around the tri-coordinated oxygen atoms and improves ferroelectricity.

## II. METHODS

The calculations are performed using first-principles density functional theory (DFT) with the projector-augmented wave (PAW) pseudopotentials[20] as implemented in the Vienna Ab Initio Simulation Package (VASP).[21] The Perdew−Burke−Ernzerhof exchange-correlation functional[22] is used. We set the kinetic energy for the plane wave basis at 500 eV, and the convergence criterion is satisfied when the differences in total energy between two consecutive electronic and ionic steps are smaller than $10^{-6}$ eV and $10^{-5}$ eV, respectively. A 6 × 6 × 6 Monkhorst−Pack[23] mesh for the k-point grid was used for the bulk HfO$_2$.

Doped films strained in the (001) and (111) orientations are modeled using supercells containing 192 and 288 atoms, respectively, and a single k-point at (¼, ¼, ¼) is used. For doping, we substitute two Hf atoms with two yttrium (Y) atoms in a supercell and compensate for the charge deficit by removing one oxygen atom (adding one O-vacancy) that is closest to the dopant. This results in 3.125% Y-doping and 1.56% O-vacancy in the 192-atom supercell representing the (001) oriented film, and 2.08% Y-doping and 1.04% O-vacancy in the 288-atom supercell representing the (111) oriented film. We also calculated the transition stress, defined as the stress required to induce a phase transformation from the non-ferroelectric monoclinic P2$_1$/c to the ferroelectric orthorhombic Pca2$_1$ phase.

## III. RESULTS AND DISCUSSION

The present study compares the monoclinic P2$_1$/c phase and the orthorhombic Pca2$_1$ phase, as the former represents the ground state under ambient conditions while the latter is of interest for its ferroelectric properties. Both phases exhibit similar structures, comprising Hf atoms in a sevenfold-coordinated configuration, along with tri-coordinated (OI) and tetra-coordinated (OII) oxygen atoms, indicated by the black and red colors in Figure 1. The spontaneous polarization in the orthorhombic Pca2$_1$ phase arises from the alignment of the OI atoms. The optimized lattice parameters and bulk modulus values are shown in Table I and agree well with previous experiments and calculations.

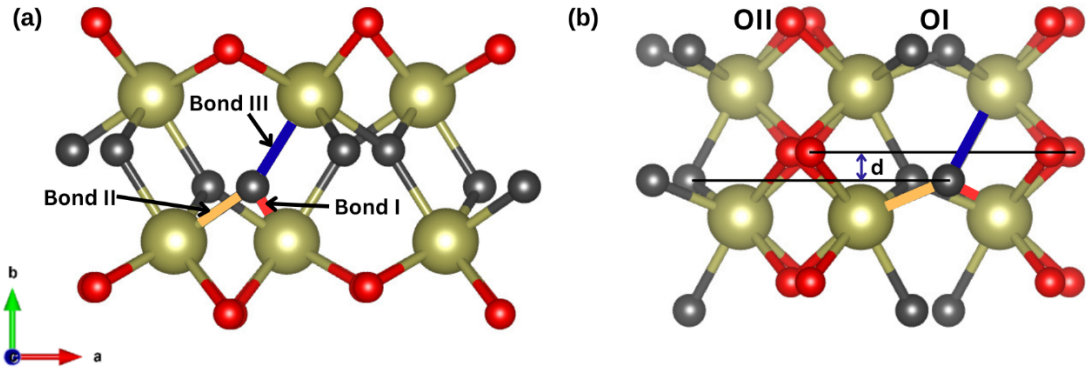

Figure 1: Structural representations of HfO$_2$ as seen along the c-axis: (a) monoclinic P2$_1$/c phase, (b) ferroelectric orthorhombic Pca2$_1$ phase. The Hf, OI, and OII atoms are shown in tan, black, and red colors, respectively. The Hf-OI bonds are labeled as bond I (red), bond II (yellow), and bond III (blue).

Table I: Calculated structural parameters for the monoclinic (P2$_1$/c) and orthorhombic (Pca2$_1$) phases of HfO$_2$ with comparison with literature.

| Phase | a (Å) | b (Å) | c (Å) | Volume (Å$^3$/ HfO$_2$) | Bulk Modulus (GPa) | Method |
|---|---|---|---|---|---|---|
|  | 5.15 | 5.20 | 5.33 | 35.15 | - | Theory[11] |
|  | 5.11 | 5.17 | 5.29 | 35.02 | 233 | Theory[24] |
| P2$_1$/c | - | - | - | 34.57 | 185 ±20 | Experiment[25] |
|  | - | - | - | 34.91 | 284 ± 10 | Experiment[26] |
|  | 5.14 | 5.18 | 5.23 | 34.94 | 242.3 | This study |
| Pca2$_1$ | 5.06 | 5.09 | 5.27 | 33.90 | - | Theory[11] |
| | 5.04 | 5.06 | 5.26 | 33.60 | 288.5 | This study |

Firstly, we show the energies of the two phases under biaxial strains in the (001) and (111) orientations in Figure 2. The four rows show the cases of pure, O-vacancy-doped, Y-doped, and co-doped HfO$_2$. In all cases, the compressive strain promotes phase transition from the monoclinic P2$_1$/c phase to the orthorhombic Pca2$_1$ phases. Interestingly, for the (111) strain, the minima of the orthorhombic Pca2$_1$ phases fall outside of the curves representing the monoclinic P2$_1$/c phase. In contrast, in the (001) strain direction, the minima of the orthorhombic Pca2$_1$ phases are found within the curves representing the monoclinic P2$_1$/c phase. These results have potential implications for the stability of the ferroelectric films. Although both the (001) and (111) strains can enable the ferroelectric phase when the area is reduced below the transition point, in case the strain is reduced due to inhomogeneity or external factors, the ferroelectric phase in films strained in the (001) orientation will likely relax by expanding and could potentially revert to the non-ferroelectric monoclinic phase, while the films strained in the (111) orientation will likely relax by contracting and remain ferroelectric. Therefore, the ferroelectricity in films strained in (111) orientation will likely be more robust than those strained in (001) orientation.

The relative energy and the area at the transition point are labeled in each panel of Figure 2. For the same type of strain, a small relative energy or a large area at the transition point would indicate an easier transition. Using these quantities, we evaluate the Y doping, O-vacancy doping, and Y and O-vacancy co-doping on the strain-induced phase transition. For the (001) orientation, the doping shows mixed effects. In all three doping scenarios, the transition energy is increased, while the transition area is increased in the co-doping case. Meanwhile, for the (111) orientation, all three doping scenarios show consistent reductions in energy and increases in the area. The most favorable case is the Y and O-vacancy co-doping, which has a transition energy of 33.20 meV/atom, compared to 35.91 meV/atom in the pure HfO$_2$ and a transition area of 11.82 Å$^2$ per Hf atom, compared to the 11.76 Å$^2$ per Hf atom in pure HfO$_2$.

To further evaluate the effects of these extrinsic factors on phase stability, we calculated the stress of transition $\sigma_\tau$ from the slope in Figure 2. As shown in Table II, the magnitudes of $\sigma_\tau$ are significant in the films strained in (001) orientation. In contrast, the magnitudes of $\sigma_\tau$ in the (111) orientation are much smaller. The case of Y and O-vacancy co-doping exhibits the lowest transition stress value among all considered cases. A reduced transition stress promotes the formation of the ferroelectric phase. Additionally, it renders the resulting film less prone to instability or degradation, thereby potentially reducing current leakage and polarization fatigue.

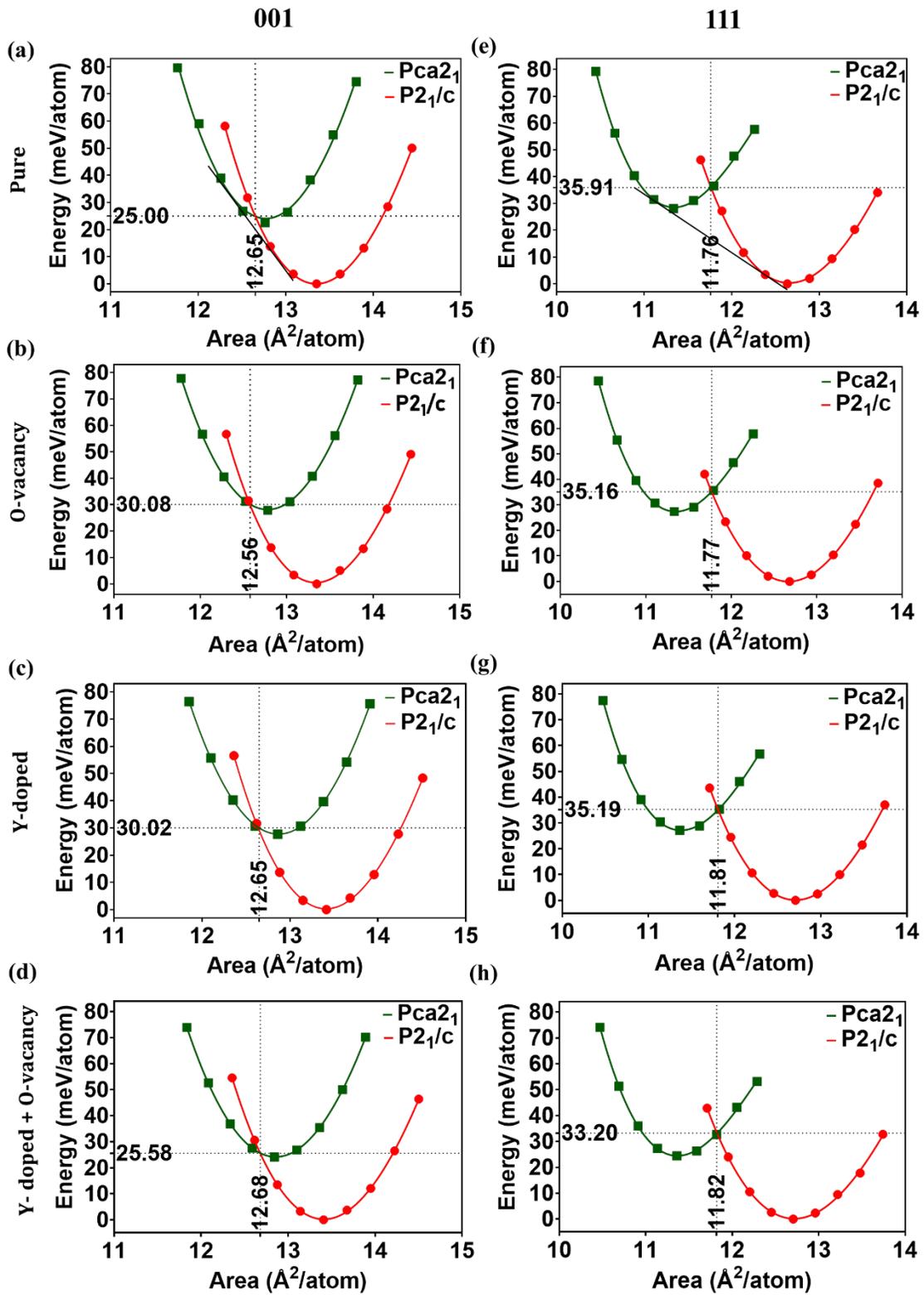

Figure 2: The relative energies of the monoclinic P2$_1$/c and orthorhombic Pca2$_1$ phases of HfO$_2$ under biaxial strains in the (001) and (111) orientations under different conditions. The energies are measured from the equilibrium energy of the monoclinic P2$_1$/c phase. The area is measured for each Hf atom.

Table II: Stress of Transition, $\sigma_\tau$ (meV/Å$^3$) under the influence of extrinsic factors

| Strain Orientation | Pure | O-vacancy | Y-doped | Y + O-vacancy |
|---|---|---|---|---|
| 001 | -49.78 | -59.09 | -57.10 | -50.16 |
| 111 | -23.97 | -23.23 | -23.37 | -19.99 |

The observations above raise three questions: (1) Why does the compressive biaxial strain favor the ferroelectric phase? (2) Why is the strain in the (111) orientation more effective? And (3) How does doping contribute to stabilization in conjunction with strain? To answer these questions, we examine the atoms responsible for the ferroelectric behavior in HfO$_2$ in more detail. As shown in Figure 1, the ferroelectricity in HfO$_2$ originates from the alignment of OI atoms with three Hf-OI bonds. (Generally, if the distance between Hf and O atoms is smaller than the sum of their atomic radii, then the Hf-O bonds are coordinated.[27–29] The atomic radii of Hf, Y, and O are 1.75, 1.90, and 0.66 Å, respectively,[30] resulting in bond length cutoffs of 2.41 Å for Hf-O and 2.56 Å for Y-O.) Figure 3 shows the variations of the bond lengths in pure-HfO$_2$ as a function of biaxial strain in both (001) and (111) orientations. It can be clearly seen that in both phases, bonds I and II decrease with increasing magnitude of compressive strain. However, bond III increases with the increasing magnitude of compressive strain in the orthorhombic Pca2$_1$ phase, while its change in the monoclinic P2$_1$/c phase is minimal. These results suggest that the distortion of bond III is crucial for understanding the emergence of ferroelectricity in HfO$_2$ under strain. The elongation of this bond under compressive strain is intriguing and indicates a more flexible local environment around OI atoms in the orthorhombic Pca2$_1$ phase compared to the monoclinic P2$_1$/c phase.

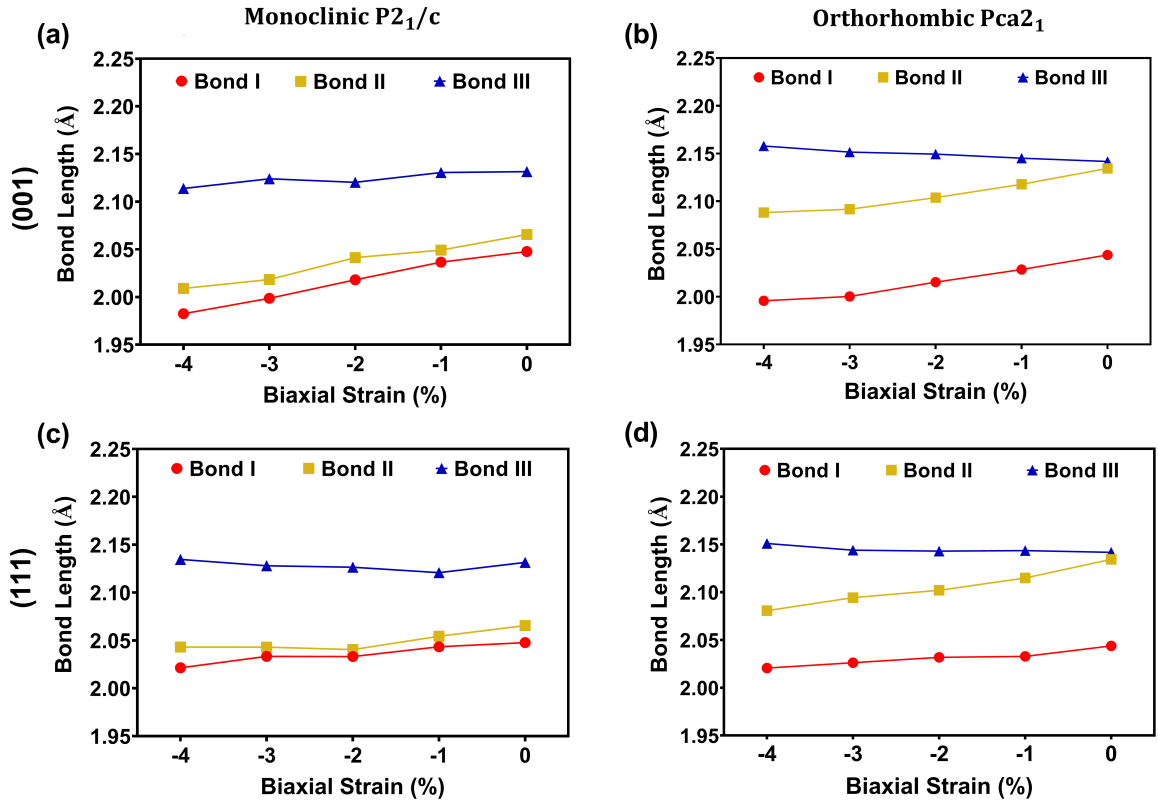

Figure 3: Lengths of Hf-OI bonds under biaxial strains in pure $HfO_2$. The variation of the bond lengths of the monoclinic $P2_1/c$ phase in the (a) (001) and (c) (111) orientation. The unique behavior of bond III is best illustrated in the orthorhombic $Pca2_1$ phase both in the (b) (001) and (d) (111) orientations. Bond I, Bond II, and Bond III are depicted in Figure 1 as red, yellow, and blue colors, respectively.

To better understand the origin of the flexibility around the tri-coordinated OI atoms, we investigate the tetra-coordinated oxygen atoms (OII) and their associated bonds (Hf-OII). Figure 4 shows the four Hf-OII bond lengths (bonds A - D) as a function of compressive (001) strain in both the monoclinic $P2_1/c$ and orthorhombic $Pca2_1$ phases. In both phases, the four bonds shorten as the compressive strain increases. However, the average length for Hf-OII bonds is shorter in the orthorhombic $Pca2_1$ phase than in the monoclinic $P2_1/c$ phase (2.204 vs 2.187 Å at 0% strain). In addition, the average bond length decreases faster in the orthorhombic $Pca2_1$ phase than in the monoclinic $P2_1/c$ phase (by -0.064 vs -0.059 Å at -4% strain). This suggests that in the orthorhombic $Pca2_1$ phase, the atomic coordination around the tetra-coordinated oxygen is not only more compact but also more compressible. Such efficient space utilization around the tetra-coordinated OII atoms leaves more room for the bonds around tri-coordinated OI atoms (it can be seen from Figure 2 that the average length of Hf-OI bonds is longer in the orthorhombic $Pca2_1$ phase), which allows easier adjustment to the compressive strain. Therefore, the compressive strain drives the energy competition in favor of the orthorhombic Pca2$_1$ phase.

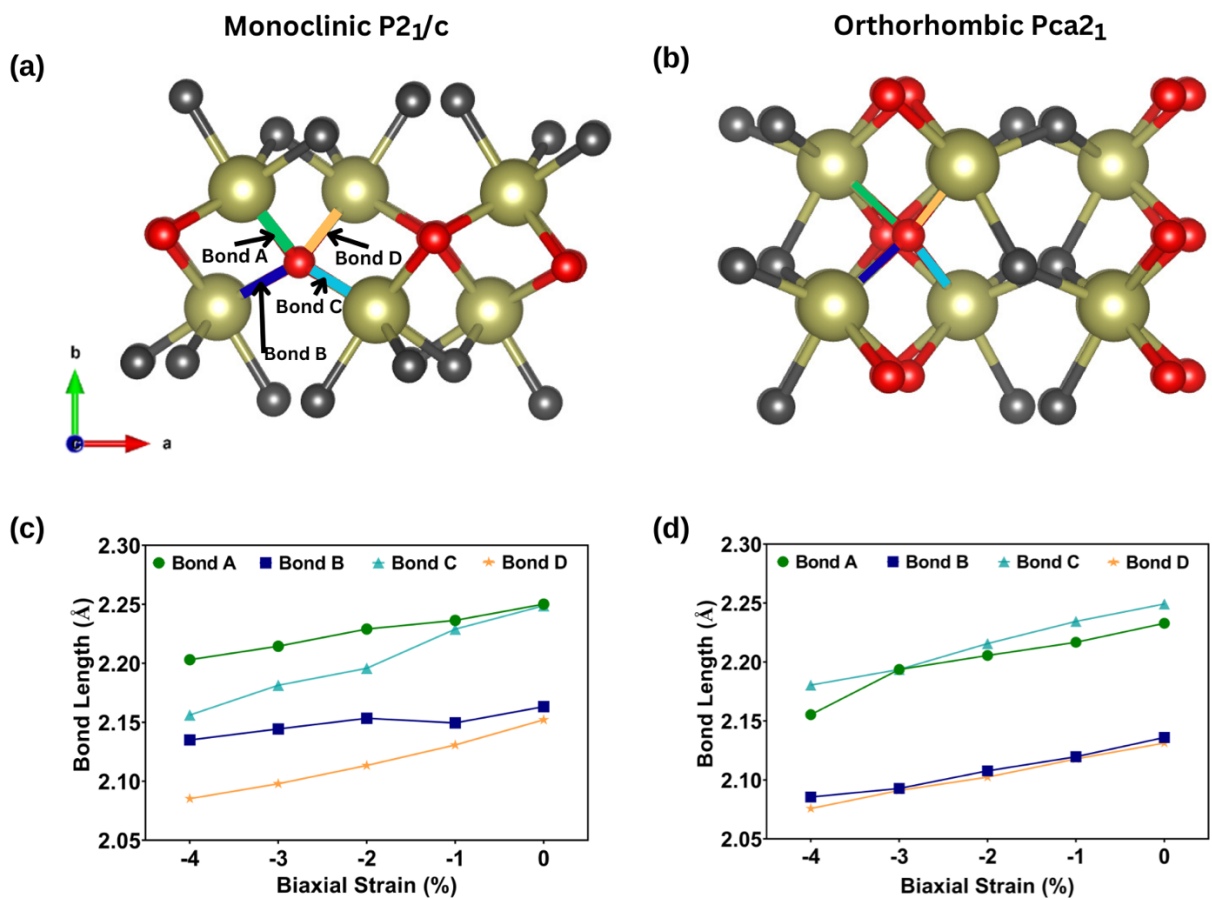

Figure 4: Structural representations and bond lengths of the hafnium atoms and tetra-coordinated oxygen (Hf-OII) under biaxial compressive strains in the (001) orientation. The Hf-OII environment is described in the (a) monoclinic P2$_1$/c and (b) orthorhombic Pca2$_1$ structures. The variations of the bond lengths are shown in (c) and (d) for the two competing phases. Bond A, bond B, bond C, and D are described by green, blue, cyan, and yellow colors, respectively.

Next, we investigate why the biaxial strain in the (111) plane is more effective than the (001) plane in stabilizing the ferroelectric orthorhombic phase. In Figures 5a and 5b, we show the orientations of these two planes relative to the orthorhombic Pca2$_1$ unit cell. The cross-section view is shown in Figures 5c and 5d. The Hf atoms form a square lattice in the (001) plane and a hexagonal lattice (111) plane. The corresponding area densities are 7.8 Hf/nm$^2$ and 8.8 Hf/nm$^2$, respectively. Due to the higher area density of Hf in the (111) plane, the biaxial compressive strain will result in a stronger out-of-plane response. In addition, the placement of oxygen atoms may also play a role. Figures 5e and 5f show the side views of the two planes. While the oxygen atoms are situated off the plane of Hf atoms in the (001) case, a fraction of OI atoms (highlighted by the magenta circles) are situated within the (111) plane of Hf atoms. Such arrangement further limits the in-plane adjustably and enhances the effect of compressive strain.

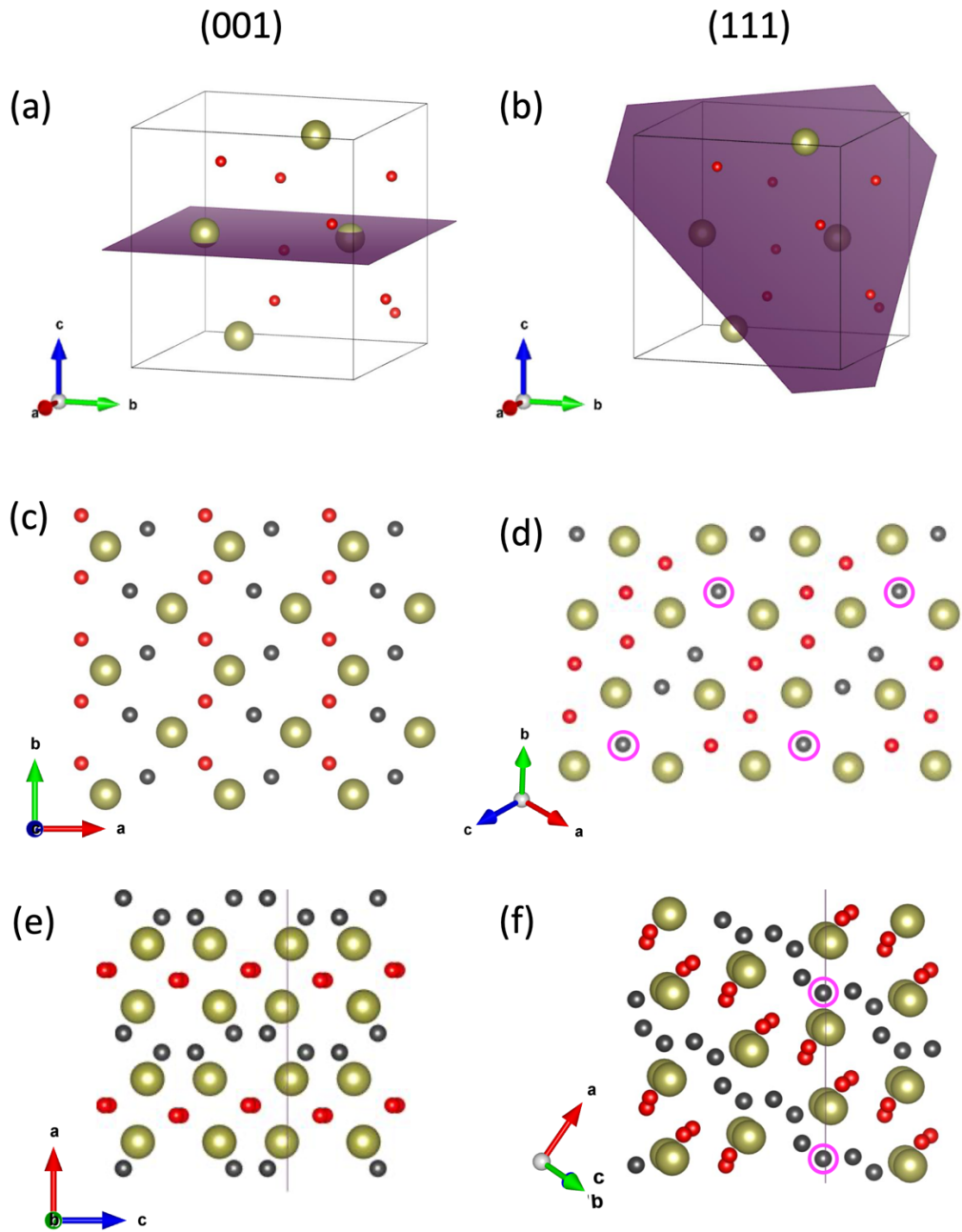

Figure 5: Crystallographic illustration of the orthorhombic Pca2$_1$ phase of HfO$_2$. Panels (a) and (b) show the (001) and (111) planes in the unit cell. Panels (c) and (d) show cross-sections of the (001) and (111) planes. Panels (e) and (f) are the side views of the (001) and (111) planes.

The remaining question is how the doping contributes to the stabilization of the ferroelectric phase in conjunction to the strain. In general, the introduction of dopants such as yttrium and/or O-vacancy alters the local bonding environment, thereby creating local strains that can enhance the effect of the external strain. Here, we propose an additional mechanism. As discussed previously, compared to the monoclinic phase, the average length of Hf-OI bonds is longer in the ferroelectric orthorhombic phase, indicating a more flexible local environment around OI atoms that can respond more favorably to a compressive strain. In Table III, we show the three Hf-OI bonds in the ferroelectric orthorhombic phase at -3% biaxial strain for pure $HfO_2$ and with Y and O-vacancy co-doping. For both (001) and (111) strain orientation, the co-doping clearly resulted in an increase in the average length of Hf-OI bonds, indicating it makes the local environment around OI atoms more flexible, which in turn makes the ferroelectric phase more favorable under compressive strain.

Table III: Bond lengths (in Å) under the influence of extrinsic factors at biaxial strain of -3%

| Strain Orientation | Extrinsic Factor | Bond I | Bond II | Bond III | Average |
|---|---|---|---|---|---|
| 001 | Pure | 2.002 | 2.064 | 2.142 | 2.069 |
|  | Y + O-Vacancy | 2.032 | 2.070 | 2.148 | 2.083 |
| 111 | Pure | 2.020 | 2.085 | 2.139 | 2.081 |
|  | Y + O-Vacancy | 2.090 | 2.119 | 2.126 | 2.112 |

Finally, to evaluate how the strain and doping affect the magnitude of ferroelectric polarization, we calculate ferroelectric displacement, shown as "d" in Figure 1, for a strain value of -3%. As shown in Table IV, the ferroelectric displacement under (111) strain orientation is higher than those in the (001) orientation, except for the O-vacancy doped case. The largest ferroelectric displacement appears in the co-doping case under the (111) strain orientation, suggesting that this combination of strain and doping is not only good for phase stability but also has the strongest polarization that is beneficial for applications.

Table IV: Ferroelectric displacement (in Å) under the influence of extrinsic factors

| Strain Orientation | Pure | O-vacancy | Y-doped | Y-doped + O-vacancy |
|---|---|---|---|---|
| 001 | 0.54 | 0.74 | 0.71 | 0.73 |
| 111 | 0.60 | 0.72 | 0.73 | 0.76 |

## IV. SUMMARY

In summary, we presented the atomistic mechanisms of ferroelectricity in hafnium oxide using density functional theory calculations. Our results show that the combination of Y-doping and O-vacancy is beneficial in reducing the transition stress, especially in the (111) plane, hence providing an optimal path for achieving the ferroelectric phase. An analysis reveals that the ferroelectric phase features a more flexible local environment for tri-coordinated oxygen atoms that allows more favorable responses to compressive strain. The higher atomic density in the (111) plane enhances the effectiveness of the strain in this orientation. The Y and O-vacancy co-doping improves the local flexibility around the tri-coordinated oxygen atoms and further enhances the ferroelectricity. These findings provide insights into the atomistic processes that drive phase stability, which can inform the design of $HfO_2$-based ferroelectric materials and applications.

## Acknowledgments

Computational resources were provided by The University of Memphis High-Performance Computing Center (HPCC). We thank Dr. Kai Ni for the helpful discussions.